\documentclass[aps,floats]{revtex4}
\usepackage{amsmath,amssymb}
\usepackage{graphicx,epsfig}
\usepackage[greek,english]{babel}

\begin{document}
\bibliographystyle {plain}

\def\oppropto{\mathop{\propto}} 
\def\opsimeq{\mathop{\simeq}}
\def\opoverderline{\mathop{\overline}}
\def\operarrow{\mathop{\longrightarrow}}
\def\opsim{\mathop{\sim}}

\def\fig#1#2{\includegraphics[height=#1]{#2}}
\def\figx#1#2{\includegraphics[width=#1]{#2}}


\title{  Dyson hierarchical quantum ferromagnetic Ising chain \\
with pure or random transverse fields} 


\author{ C\'ecile Monthus }
 \affiliation{Institut de Physique Th\'{e}orique, Universit\'e Paris Saclay, CNRS, CEA,
91191 Gif-sur-Yvette, France}

\begin{abstract}

The Dyson hierarchical version of the quantum Ising chain with Long-Ranged power-law ferromagnetic couplings $J(r) \propto r^{-1-\sigma}$ and pure or random transverse fields is studied via real-space renormalization. For the pure case, the critical exponents are explicitly obtained as a function of the parameter $\sigma$, and are compared with previous results of other approaches. For the random case, the RG rules are numerically applied and the critical behaviors are compared with previous Strong Disorder Renormalization results.

\end{abstract}

\maketitle

\section{ Introduction }

The quantum Ising model is the basic model in the field of zero-temperature quantum phase transitions \cite{sachdev}.
For the short-ranged case on hypercubic lattice in dimension $d$ \cite{sachdev}, 
 the dynamical exponent relating time and space via $t=L^z$ takes the simple value
\begin{eqnarray}
z^{pure}_{SR}=1
\label{zpureSR}
\end{eqnarray}
This means that the time simply plays the role of an additional spatial dimension
in the quantum-classical correspondence, so that the $d$-dimensional quantum Ising model
is equivalent to the {\it classical } Ising model in $D=d+1$ dimensions.
It is also interesting to consider the long-ranged quantum Ising chain 
\begin{eqnarray}
H^{pure}_{LR}= - \sum_{i}  h \sigma^z_i -\frac{1}{2} \sum_{i,j} J_{i,j} \sigma_i^x \sigma^x_j 
\label{hqising}
\end{eqnarray}
with uniform transverse field $h$ and power-law ferromagnetic coupling
\begin{eqnarray}
J_{i,j}= \frac{1}{\vert i-j \vert ^{1+\sigma}}
\label{jLR}
\end{eqnarray}
in order to study how the critical properties 
depend upon the parameter $\sigma>0$ \cite{dutta2001}.
This quantum chain is equivalent to a $2D$ classical 
Ising model with nearest-neighbor coupling in the 'time' direction and long-ranged
power-law coupling in the 'spatial' direction, so that the dynamical exponent $z(\sigma)$
is not given by the short-ranged isotropic value of Eq. \ref{zpureSR}
anymore, as a consequence of the strong anisotropy between time and space.
 In particular in the mean-field region $0<\sigma<\sigma_u=2/3$ \cite{dutta2001}
(see the reminder in Appendix \ref{sec_meanfield}),
the dynamical exponent $z$, the correlation length exponent $\nu$
and the anomalous dimension $\eta$ read \cite{dutta2001}
\begin{eqnarray}
z^{pure}_{MF}(0<\sigma<\sigma_u=2/3) && = \frac{\sigma}{2}
\nonumber \\
\nu^{pure}_{MF}(0<\sigma<\sigma_u=2/3) && = \frac{1}{\sigma}
\nonumber \\
\eta^{pure}_{MF}(0<\sigma<\sigma_u=2/3) && = 2-\sigma
\label{zpureLRMF}
\end{eqnarray}
It turns out that the other problem of the {\it dissipative } short-ranged
quantum spin chain
 is equivalent after integration over the bath degrees of freedom
to a $2D$ classical model with nearest-neighbor coupling in the 'spatial' direction
 and long-ranged power-law coupling in the 'time' direction,
where the parameter $\sigma$ now characterizes the bath spectral function $J(\omega \to 0) \propto \omega^{\sigma}$ \cite{werner2005,werner2005bis,norway2010,norway2012}:
the value $\sigma=1$ corresponds to the most studied Ohmic damping, 
whereas the region $0<\sigma<1$ corresponds to sub-Ohmic damping,
and the region $\sigma>1$ to super-Ohmic damping.
The dissipative short-ranged chain is thus equivalent to the model of Eq. \ref{hqising} after the interchange of space and time.
So the dynamical exponent $z'$,
the correlation length exponent $\nu'$, and the anomalous dimension $\eta'$
of the dissipative quantum chain are given by $z'=1/z$, $\nu'=\nu z$
and $z+\eta-1=\frac{z'+\eta'-1}{z'} $
As a consequence, the values $z'\simeq 2$, $\nu' \simeq 0.64$ and $\eta'=0$
measured via Monte-Carlo for the dissipative Ohmic chain \cite{werner2005,norway2010,norway2012}
translate for the quantum chain of Eq. \ref{hqising} into
\begin{eqnarray}
z^{pure}_{LR}(\sigma =1) && \simeq 0.5
\nonumber \\
\nu^{pure}_{LR}(\sigma =1) && \simeq 1.28
\nonumber \\
\eta^{pure}_{LR}(\sigma =1) && \simeq 1.
\label{zpureLRnume}
\end{eqnarray}

Let us now consider the effects of randomness in the transverse-fields $h_i$  
\begin{eqnarray}
H_{LR}^{random}= 
- \sum_{i}  h_i \sigma^z_i - \sum_{(i,j)} J_{i,j} \sigma_i^x \sigma^x_j 
\label{hqisingrandom}
\end{eqnarray}
The relevance of a small disorder at pure quantum phase transitions 
needs to be discussed from two points of view \cite{vojta} :
on one hand the Harris criterion \cite{harris}  
or equivalently the Chayes {\it et al} inequality \cite{chayes}
imply that the pure fixed point can be stable only if $\nu^{pure} \geq 2/d=2$
here in the spatial dimension $d=1$;
on the other hand, the analysis of rare regions \cite{vojta} shows that it is important
to compare the dimensionality $d_{RR}$ of rare regions (here $d_{RR}=1$
since the disorder is actually 'infinitely' correlated along the time-direction)
and the lower critical dimension $d^{class}_l=1$ sufficient to obtain 
ferromagnetic ordering : the case $d_{RR}=d^{class}_l$ corresponds to models where
rare regions can play an essential role at criticality (see \cite{vojta} for more details). The random LR chain of Eq. \ref{hqisingrandom} with a finite initial disorder
has been studied recently \cite{strongLR}
via the Strong Disorder Renormalization (see \cite{review_strong} for a review) :
the main results \cite{strongLR} (see also the related work \cite{epiLR} in arbitrary 
dimension $d$)
are the following critical dynamical exponent
\begin{eqnarray}
z^{random}_{SDRG}(\sigma ) =1+\sigma
\label{zrandomSD}
\end{eqnarray}
and the essential singularity of the correlation length 
as the control parameter $\theta$ approaches its critical value $\theta_c$
\begin{eqnarray}
\ln \xi^{random}_{SDRG} \propto \frac{1}{\vert \theta-\theta_c \vert}
\label{xirandomSD}
\end{eqnarray}
corresponding formally to an infinite correlation length exponent 
\begin{eqnarray}
\nu^{random}_{SDRG}(\sigma ) =+\infty
\label{nurandomSD}
\end{eqnarray}
These properties should be contrasted with the short-ranged random Chain
 governed by an Infinite Disorder Fixed Point characterized by
an infinite dynamical exponent 
\begin{eqnarray}
z^{random}_{SR}(\sigma ) =+\infty
\label{zSRinfty}
\end{eqnarray}
 and two finite correlation length exponents $\nu^{random}_{SRtyp}=1$ and $\nu^{random}_{SRav}=2$
\cite{fisher}.
Note that the effects of disorder on the dissipative quantum chain
mentioned above
has been also much studied via Strong Disorder RG  \cite{rieger,hoyos}
and via Monte-Carlo \cite{vojtadissi}, but here the problem is not equivalent to Eq.
\ref{hqisingrandom}, as a consequence of the columnar nature of the disorder
along the time direction in each problem (the interchange of space and time
discussed above for the pure case is not possible anymore).

Since the obtained 
critical dynamical exponent of Eq. \ref{zrandomSD} is finite and not infinite
as in the short-ranged case (Eq. \ref{zSRinfty}),
the Strong Disorder RG approach is not exact asymptotically, but only approximate.
As a consequence, it seems useful to analyze the critical properties
 with another approach in order to compare the results.
For the random Short-Ranged chain, the seld-dual block RG procedure
first introduced for the pure chain \cite{pacheco} has been found recently
to be able to reproduce the Fisher Infinite-Disorder fixed point 
\cite{nishiRandom,c_pacheco2d,c_renyi} : this shows that the same block RG
rules can lead to conventional critical behavior for the pure chain and
to Infinite disorder critical behavior for the random chain
 depending on the initial condition of the RG flow.
The aim of the present paper is to study via some block renormalization
the Dyson hierarchical analog of the pure and the random long-ranged chain
of Eq. \ref{hqising} and Eq. \ref{hqisingrandom}. 

The paper is organized as follows. 
In section \ref{sec_rgdyson}, we introduce the Dyson hierarchical quantum Ising model
and derive the renormalization rules.
In section \ref{sec_pure}, the RG equations for the Dyson analog of the
pure long-ranged chain of Eq. \ref{hqising} are solved analytically.
In section \ref{sec_random}, the RG rules for the Dyson analog of the
random long-ranged chain of Eq. \ref{hqisingrandom}
are studied numerically to obtain the critical properties.
Our conclusions are summarized in section \ref{sec_conclusion}.
 Appendix \ref{sec_meanfield} contains a reminder on the mean-field theory
for the pure long-ranged chain.

\section{ Renormalization rules for the Dyson hierarchical model }

\label{sec_rgdyson}

\subsection{ Dyson hierarchical version of the Long-Ranged Quantum Ising Chain }

In the field of long ranged models, it is very useful to consider
their Dyson hierarchical analogs, where real space renormalization 
procedures are usually easier to define and to solve
as a consequence of the hierarchical structure. 
The Dyson hierarchical classical ferromagnetic Ising model
 \cite{dyson} has been much studied by both mathematicians
\cite{bleher,gallavotti,book,jona} and physicists \cite{baker,mcguire,Kim,Kim77,us_dysonferrodyn}. More recently, Dyson hierarchical versions 
 have been considered for various disordered systems,
either classical like random fields Ising models
\cite{randomfield,us_aval} and spin-glasses \cite{franz,castel_etal,castel_parisi,castel,angelini}, or quantum like Anderson localization models \cite{bovier,molchanov,krit,kuttruf,fyodorov,EBetOG,fyodorovbis,us_dysonloc}.

Here we introduce the Dyson hierarchical analog of the Long-Ranged Quantum Ising Chain of Eq. \ref{hqisingrandom} as follows.
The Hamiltonian for $2^n$ quantum spins can be decomposed
 as a sum over the generations $k=0,1,..,n-1$
\begin{eqnarray}
H_{(1,2^n)} &&  = \sum_{k=0}^{n-1} H^{(k)}_{(1,2^n)} 
\label{recDyson}
\end{eqnarray}
The Hamiltonian of generation $k=0$ contains the 
transverse fields $h_i$ and the lowest order couplings $J^{(0)}$ 
\begin{eqnarray}
 H^{(k=0)}_{(1,2^n)} &&  = - \sum_{i=1}^{2^n} h_i \sigma_i^z
 - \sum_{i=1}^{2^{n-1}}  J^{(0)} [\mu_{2i-1}   \sigma_{2i-1}^x] [ \mu_{2i} \sigma_{2i}^x]
\label{h0dyson}
\end{eqnarray}
The Hamiltonian of generation $k=1$ reads
\begin{eqnarray}
 H^{(k=1)}_{(1,2^n)} &&  =
 - \sum_{i=1}^{2^{n-2}}  J^{(1)} 
\left[ \mu_{4i-3}   \sigma_{4i-3}^x + \mu_{4i-2} \sigma_{4i-2}^x \right]
\left[ \mu_{4i-1}   \sigma_{4i-1}^x + \mu_{4i} \sigma_{4i}^x \right]
\label{h1dyson}
\end{eqnarray}
the Hamiltonian of generation $k=2$ reads
\begin{eqnarray}
 H^{(k=2)}_{(1,2^n)}   =
 - \sum_{i=1}^{2^{n-3}}  J^{(2)} 
&& \left[ \mu_{8i-7}   \sigma_{8i-7}^x + \mu_{8i-6} \sigma_{8i-6}^x+
 \mu_{8i-5}   \sigma_{8i-5}^x + \mu_{8i-4} \sigma_{8i-4}^x \right]
\\ \nonumber 
\times && \left[ \mu_{8i-3}   \sigma_{8i-3}^x + \mu_{8i-2} \sigma_{8i-2}^x+
 \mu_{8i-1}   \sigma_{8i-1}^x + \mu_{8i} \sigma_{8i}^x \right]
\label{h2dyson}
\end{eqnarray}
and so on up to the last generation $k=n-1$ that couples the two halves of the system
\begin{eqnarray}
 H^{(n-1)}_{(1,2^n)} = 
- J^{(n-1)}  \left[ \sum_{i=1}^{2^{n-1}} \mu_i \sigma_i^x  \right]
\left[ \sum_{j=2^{n-1}+1}^{2^n}  \mu_j  \sigma_j^x \right]
\label{hlastdyson}
\end{eqnarray}

The transverse fields $h_i>0$ in Eq. \ref{h0dyson}
can be either uniform or random.
The magnetic moments $\mu_i$ of the spins are set initially to unity
\begin{eqnarray}
\mu_i=1
\label{muiini}
\end{eqnarray}
but we have introduced them because they will be generated by the renormalization procedure described below.

The $J^{(k)}>0$ are given ferromagnetic couplings as a function of the generation $k$.
 To mimic the power-law behavior with respect to the distance $r$
of Eq. \ref{jLR}
\begin{eqnarray}
J(r) = \frac{1}{r^{1+\sigma}}
\label{powerr}
\end{eqnarray}
we consider the following exponential behavior with respect
to the generation $k$
\begin{eqnarray}
J^{(k)} = \frac{1}{(2^k)^{1+\sigma}}= 2^{-(1+\sigma)k}
\label{powerk}
\end{eqnarray}

At the classical level, the energy cost with respect to the ground state
of a Domain-Wall
between the first half-system having $S^x=1$ for $1 \leq i \leq \frac{L}{2}$
and the second half-system having $S^x=-1$ for $\frac{L}{2}+1 \leq i \leq L$
scales as
\begin{eqnarray}
E^{DW}(L) \propto L^{1-\sigma} 
\label{edw}
\end{eqnarray}
In the region $\sigma>1$ where it decays with the system-size $L$,
it is clearly different from the Long-Ranged model of Eq. \ref{hqising}
that at least contains a constant term coming from the nearest-neighbor coupling
between the two halves. On the contrary in the region $\sigma<1$ where 
Eq. \ref{edw} grows with the system-size $L$, one may expect that the Dyson hierarchical version is an appropriate approximation of the Long-Ranged model.
The case $\sigma=1$ is at the border line, since the Domain-Wall cost
remains constant for the Dyson model (Eq. \ref{edw}), whereas it 
grows logarithmically in $L$ for the Long-Ranged model.
In the following, we will thus focus on the interval 
\begin{eqnarray}
0<\sigma \leq 1
\label{domain}
\end{eqnarray}
In particular we should stress that whereas the limit $\sigma \to + \infty$
of the Long-Ranged model of Eqs \ref{hqising} and \ref{hqisingrandom}
yields the corresponding Short-Ranged models, 
 the limitation $\sigma \leq 1$ for the Dyson hierarchical versions
does not allow to recover the Short-Ranged models in this limit.

\subsection{ Diagonalization of the lowest generation Hamiltonian $H^{(k=0)}$}

The Hamiltonian $H^{(k=0)}$ of generation $k=0$ of Eq. \ref{h0dyson}
is the sum of the independent two-spin Hamiltonians
\begin{eqnarray}
H_{(2i-1,2i)} &&  = - h_{2i-1} \sigma_{2i-1}^z- h_{2i} \sigma_{2i}^z
- J^{(0)} \mu_{2i-1}   \mu_{2i}  \sigma_{2i-1}^x \sigma_{2i}^x
\label{h1box}
\end{eqnarray}

\subsubsection{ Diagonalization in the symmetric sector }

Within the symmetric sector, the diagonalization of the Hamiltonian of Eq. \ref{h1box}
in the $\sigma^z$ basis
\begin{eqnarray}
H_{(2i-1,2i)} \vert ++> && = - (h_{2i-1}+h_{2i}) \vert ++ >-  J^{(0)}\mu_{2i-1}   \mu_{2i} \vert -- >
\nonumber \\
H_{(2i-1,2i)} \vert --> && =-  J^{(0)} \mu_{2i-1}   \mu_{2i}\vert ++ >  -(h_{2i-1}+h_{2i})  \vert -- >
\label{huvs}
\end{eqnarray}
leads to the two eigenvalues
\begin{eqnarray}
\lambda_{2i}^{S-} = - \sqrt{ (J^{(0)}\mu_{2i-1}   \mu_{2i})^2+(h_{2i-1}+h_{2i})^2 }
\nonumber \\
\lambda_{2i}^{S+} = + \sqrt{ (J^{(0)}\mu_{2i-1}   \mu_{2i})^2+(h_{2i-1}+h_{2i})^2 }
\label{lambdas}
\end{eqnarray}
with the corresponding eigenvectors 
\begin{eqnarray}
\vert \lambda_{2i}^{S-} > && = \cos \theta_{2i}^S\vert ++>+\sin \theta_{2i}^S\vert -- >
\nonumber \\
\vert \lambda_{2i}^{S+} > && = -\sin \theta_{2i}^S\vert ++ > + \cos \theta_{2i}^S\vert -->
\label{vlambdas}
\end{eqnarray}
in terms of the angle $\theta_S$ satisfying
\begin{eqnarray}
\cos ( \theta_{2i}^S ) && = 
\sqrt{ \frac{1+ \frac{h_{2i-1}+h_{2i}}{ \sqrt{ (J^{(0)} \mu_{2i-1}   \mu_{2i})^2+(h_{2i-1}+h_{2i})^2 }}}{2}}
\nonumber \\
\sin ( \theta_{2i}^S ) && =
\sqrt{ \frac{1- \frac{h_{2i-1}+h_{2i}}{ \sqrt{ (J^{(0)} \mu_{2i-1}   \mu_{2i})^2+(h_{2i-1}+h_{2i})^2 }}}{2}}
\label{thetas}
\end{eqnarray}

\subsubsection{Diagonalization in the antisymmetric sector  }

Within the antisymmetric sector
\begin{eqnarray}
H^{(1)}_{(2i-1,2i)} \vert +-> && = - (h_{2i-1}-h_{2i}) \vert +- >-  J^{(0)}\mu_{2i-1}   \mu_{2i} \vert -+ >
\nonumber \\
H^{(1)}_{(2i-1,2i)} \vert -+> && =-  J^{(0)}\mu_{2i-1}   \mu_{2i} \vert +- >  +(h_{2i-1}-h_{2i})  \vert -+ >
\label{huva}
\end{eqnarray}
 the two eigenvalues read
\begin{eqnarray}
\lambda_{2i}^{A-} = - \sqrt{ (J^{(0)} \mu_{2i-1}   \mu_{2i})^2+(h_{2i-1}-h_{2i})^2 }
\nonumber \\
\lambda_{2i}^{A+} = + \sqrt{ (J^{(0)} \mu_{2i-1}   \mu_{2i})^2+(h_{2i-1}-h_{2i})^2 }
\label{lambdaa}
\end{eqnarray}
with the corresponding eigenvectors 
\begin{eqnarray}
\vert \lambda_{2i}^{A-} > && = \cos \theta_{2i}^A\vert +->+\sin \theta_{2i}^A\vert -+ >
\nonumber \\
\vert \lambda_{2i}^{A+} > && = -\sin \theta_{2i}^A\vert +- > + \cos \theta_{2i}^A\vert -+>
\label{vlambdaa}
\end{eqnarray}
in terms of the angle $\theta_A$ satisfying
\begin{eqnarray}
\cos (  \theta_{2i}^A ) && =
\sqrt{ \frac{1+ \frac{h_{2i-1}-h_{2i}}{ \sqrt{ (J^{(0)} \mu_{2i-1}   \mu_{2i})^2+(h_{2i-1}-h_{2i})^2 }}
 }{2}}
\nonumber \\
\sin (  \theta_{2i}^A ) && =
\sqrt{ \frac{1- \frac{h_{2i-1}-h_{2i}}{ \sqrt{ (J^{(0)} \mu_{2i-1}   \mu_{2i})^2+(h_{2i-1}-h_{2i})^2 }}
 }{2}}
\label{thetaa}
\end{eqnarray}

\subsection{ Introduction of the renormalized spins $\sigma_{R(2i)}$}

For each two-spin Hamiltonian $H_{2i-1,2i}$ of Eq. \ref{h1box},
we wish to keep the two lowest states among the four eigenstates discussed
 above, and to label them as the two states
of some renormalized spin $\sigma_{R(2i)}$
\begin{eqnarray}
\vert \sigma^z_{R(2i)}=+>  && \equiv \vert \lambda_{2i}^{S-} > 
\nonumber \\ 
\vert \sigma^z_{R(2i)}=->  && \equiv \vert \lambda_{2i}^{A-} > 
\label{sigmaR2states}
\end{eqnarray}
It is convenient to introduce the corresponding projector 
\begin{eqnarray}
P_{2i}^- \equiv \vert \sigma^z_{R(2i)}=+ > < \vert \sigma^z_{R(2i)}=+ \vert  + 
\vert \sigma^z_{R(2i)}=- > < \sigma^z_{R(2i)}=- \vert
\label{proj}
\end{eqnarray}
as well as the spin operators
\begin{eqnarray}
\sigma^z_{R(2i)} && \equiv \vert \sigma^z_{R(2i)}=+ > < \vert \sigma^z_{R(2i)}=+ \vert  - 
\vert \sigma^z_{R(2i)}=- > < \sigma^z_{R(2i)}=- \vert
\nonumber \\ 
\sigma^x_{R(2i)} && \equiv \vert \sigma^z_{R(2i)}=+ > < \vert \sigma^z_{R(2i)}=- \vert  + 
\vert \sigma^z_{R(2i)}=- > < \sigma^z_{R(2i)}=+ \vert
\label{opsigmaR}
\end{eqnarray}

\subsection{ Renormalization rule for the transverse fields $ h_{R(2i)} $}

The projection of the Hamiltonian of Eq. \ref{h1box} is given by
\begin{eqnarray}
P_{2i}^- H_{(2i-1,2i)} P_{2i}^- && =
 \lambda_{2i}^{S-} \vert \lambda_{2i}^{S-} > <\lambda_{2i}^{S-} \vert
 + \lambda_{2i}^{A-} \vert \lambda_{2i}^{A-} > <\lambda_{2i}^{A-} \vert
\nonumber \\
&& =\left( \frac{ \lambda_{2i}^{S-}+\lambda_{2i}^{A-}  }{2} \right)  P_{2i}^- 
  + \left( \frac{ \lambda_{2i}^{S-}-\lambda_{2i}^{A-}  }{2} \right)  \sigma^z_{R(2i)}
\nonumber \\
&& \equiv  e_{R(2i)}  P_{2i}^-  - h_{R(2i)} \sigma^z_{R(2i)}
\label{projH}
\end{eqnarray}
where the renormalized transverse fields read
\begin{eqnarray}
 h_{R(2i)} && \equiv \frac{ \lambda_{2i}^{A-}-\lambda_{2i}^{S-}  }{2}
\nonumber \\
&& = \frac{ \sqrt{ (J^{(0)}\mu_{2i-1}   \mu_{2i} )^2+(h_{2i-1}+h_{2i})^2 }
- \sqrt{ (J^{(0)}\mu_{2i-1}   \mu_{2i})^2+(h_{2i-1}-h_{2i})^2 }  }{2}
\nonumber \\
&& = \frac{ 2 h_{2i-1} h_{2i} }
{\sqrt{ (J^{(0)}\mu_{2i-1}   \mu_{2i} )^2+(h_{2i-1}+h_{2i})^2 }
+ \sqrt{ (J^{(0)}\mu_{2i-1}   \mu_{2i})^2+(h_{2i-1}-h_{2i})^2 }}
\label{rgh}
\end{eqnarray}
and where the contribution to the ground-state energy of this projection reads
\begin{eqnarray}
 e_{R(2i)} && \equiv \frac{ \lambda_{2i}^{A-}+\lambda_{2i}^{S-}  }{2}
\nonumber \\
&& =- \frac{ \sqrt{ (J^{(0)}\mu_{2i-1}   \mu_{2i} )^2+(h_{2i-1}+h_{2i})^2 }
+ \sqrt{ (J^{(0)}\mu_{2i-1}   \mu_{2i})^2+(h_{2i-1}-h_{2i})^2 }  }{2}
\label{rgegs}
\end{eqnarray}

\subsection{ Renormalization rule for the magnetic moments $\mu_{R(2i)}$}

The projections of the $\sigma^x$ operators
\begin{eqnarray}
P_{2i}^- \sigma^x_{2i-1} P_{2i}^- && =
\left[\sin (\theta_{2i}^S) \cos (\theta_{2i}^A)+ \cos (\theta_{2i}^S) \sin (\theta_{2i}^A)  \right] \sigma^x_{R(2i)}
\nonumber \\
P_{2i}^- \sigma^x_{2i} P_{2i}^- && =
\left[\cos (\theta_{2i}^S) \cos (\theta_{2i}^A)+ \sin (\theta_{2i}^S) \sin (\theta_{2i}^A)  \right]\sigma^x_{R(2i)}
\label{newbasis}
\end{eqnarray}
yield the following renormalization rule for the magnetic moment
\begin{eqnarray}
\mu_{R(2i)} && = \left[\sin (\theta_{2i}^S) \cos (\theta_{2i}^A)+ \cos (\theta_{2i}^S) \sin (\theta_{2i}^A)  \right] \mu_{2i-1}+
\left[\cos (\theta_{2i}^S) \cos (\theta_{2i}^A)+ \sin (\theta_{2i}^S) \sin (\theta_{2i}^A)  \right]  \mu_{R(2i)}
\nonumber \\
&& =\mu_{2i-1}  \sqrt{ \frac{1+\frac{(J^{(0)} \mu_{2i-1}   \mu_{2i})^2-h_{2i-1}^2+h_{2i}^2 }{\sqrt{ (J^{(0)} \mu_{2i-1}   \mu_{2i})^2+(h_{2i-1}+h_{2i})^2 }
\sqrt{ (J^{(0)} \mu_{2i-1}   \mu_{2i})^2+(h_{2i-1}-h_{2i})^2 }}}{2}  }
\nonumber \\
&& +  \mu_{2i} \sqrt{ \frac{1+\frac{(J^{(0)} \mu_{2i-1}   \mu_{2i})^2+h_{2i-1}^2-h_{2i}^2 }{\sqrt{ (J^{(0)} \mu_{2i-1}   \mu_{2i})^2+(h_{2i-1}+h_{2i})^2 }
\sqrt{ (J^{(0)} \mu_{2i-1}   \mu_{2i})^2+(h_{2i-1}-h_{2i})^2 }}}{2}  }
\label{RGmu}
\end{eqnarray}

\subsection{ Renormalization of the Hamiltonians of generation $k \geq 1$}

The Hamiltonian of generation $k=1$ of Eq. \ref{h1dyson}
is projected onto
\begin{eqnarray}
\left( \prod_{i=1}^{2^{n-1}} P_{2i}^- \right) H^{(k=1)}_{(1,2^n)} 
\left( \prod_{i=1}^{2^{n-1}} P_{2i}^- \right) &&  =
 - \sum_{i=1}^{2^{n-2}}  J^{(1)} 
\left[\mu_{R(4i-2)} \sigma_{R(4i-2)}^x \right]
\left[ \mu_{R(4i)} \sigma_{R(4i)}^x \right]
\label{h1dysonrg}
\end{eqnarray}
The Hamiltonian of generation $k=2$ of Eq. \ref{h2dyson} 
is projected onto
\begin{eqnarray}
\left( \prod_{i=1}^{2^{n-1}} P_{2i}^- \right) H^{(k=2)}_{(1,2^n)} \left( \prod_{i=1}^{2^{n-1}} P_{2i}^- \right)  =
 - \sum_{i=1}^{2^{n-3}}  J^{(2)} 
\left[  \mu_{R(8i-6)} \sigma_{R(8i-6)}^x+
  \mu_{R(8i-4)} \sigma_{R(8i-4)}^x \right]
 \left[ \mu_{R(8i-2)} \sigma_{R(8i-2)}^x+
 + \mu_{R(8i)} \sigma_{R(8i)}^x \right]
\label{h2dysonrg}
\end{eqnarray}
and so on up to the Hamiltonian of last generation $k=n-1$
of Eq. \ref{hlastdyson} that is projected onto
\begin{eqnarray}
\left( \prod_{i=1}^{2^{n-1}} P_{2i}^- \right) H^{(n-1)}_{(1,2^n)}
\left( \prod_{i=1}^{2^{n-1}} P_{2i}^- \right) = 
- J^{(n-1)}  \left[ \sum_{i=1}^{2^{n-2}} \mu_{R(2i)} \sigma_{R(2i)}^x  \right]
\left[ \sum_{j=2^{n-2}+1}^{2^{n-1}}  \mu_{R(2j)} \sigma_{R(2j)}^x \right]
\label{hlastdysonrg}
\end{eqnarray}

In conclusion, once the renormalization of the magnetic moments 
has been taken into account, the couplings $J^{(k)}$ 
are just translated by one generation for $k \geq 0$
\begin{eqnarray}
J^{R(k)} && = J^{(k+1)} 
\label{jnpurdyson}
\end{eqnarray}

\section{ Dyson hierarchical chain with uniform transverse field }

\label{sec_pure}

The pure Dyson Quantum ferromagnetic Ising model corresponds
 to the case where all transverse fields $h_i$ coincide
\begin{eqnarray}
h_i  = h 
\label{h0pur}
\end{eqnarray}
The mean-field theory for the Long-Ranged model of Eq. \ref{hqising}
 \cite{dutta2001} (see the reminder in Appendix \ref{sec_meanfield}) 
 is expected to be also valid for the Dyson hierarchical
version in the same region $0<\sigma<\sigma_u=2/3$.
The main goal of the real-space procedure described below 
is to study the critical properties in the non-mean-field region $\sigma_u \geq 2/3$,
but since $\sigma$ is just a continuous parameter in the RG rules,
we will also mention the results for $0<\sigma<\sigma_u=2/3 $.

\subsection{ Pure renormalization rules }

The renormalization rules derived in the previous section simplify as follows :

(i) the renormalized transverse field of Eq. \ref{rgh} reads
\begin{eqnarray}
h^R  =   \frac{ 2 h^2 }
{J^{(0)} \mu^2 + \sqrt{ (J^{(0)}\mu^2 )^2+4 h^2 } }
\label{hrpur}
\end{eqnarray}

(ii) the renormalized magnetic moment of Eq. \ref{RGmu} reads
\begin{eqnarray}
\mu_{R} 
&& =\mu \sqrt{2}   \sqrt{ 1+\frac{J^{(0)} \mu^2 }{\sqrt{ (J^{(0)} \mu^2)^2+4h^2 }
} }
\label{RGmupur}
\end{eqnarray}

\subsection{ RG rules deep in the paramagnetic phase $h \to +\infty$  }

Deep in the paramagnetic phase $h \to +\infty$,
the transverse field remains unchanged upon RG (Eq. \ref{hrpur})
\begin{eqnarray}
h^R  \opsimeq_{h \to + \infty}  h
\label{hrpurpara}
\end{eqnarray}
whereas the magnetic moment evolves according to (Eq. \ref{RGmupur})
\begin{eqnarray}
\mu_{R} && \opsimeq_{h \to + \infty} \mu \sqrt{2} 
\label{RGmupurpara}
\end{eqnarray}
For a length $L=2^n$ obtained after $n$ RG steps,
 the magnetic moment reads
\begin{eqnarray}
\mu(L=2^n) && \opsimeq_{h \to + \infty} (\sqrt{2})^n = \sqrt{L}
\label{muLpurpara}
\end{eqnarray}
As a consequence,
 the effective ferromagnetic coupling between two such magnetic moments
scaling as
\begin{eqnarray}
J^{(n)}_{eff} \equiv J^{(n)} \mu(L=2^n) \mu(L=2^n) && \opsimeq_{h \to + \infty} L^{- (1+\sigma)} L  = L^{-\sigma}
\label{jmumuLpurpara}
\end{eqnarray}
becomes smaller and smaller with respect to the transverse field of Eq. \ref{hrpurpara}, so that the paramagnetic fixed point is attractive for all $\sigma>0$.

\subsection{ RG rules deep in the ferromagnetic phase $h \to 0$  }

Deep in the ferromagnetic phase $h \to 0$,
 the magnetic moment evolves according to (Eq. \ref{RGmupur})
\begin{eqnarray}
\mu_{R} && \opsimeq_{h \to 0} 2 \mu 
\label{RGmupurferro}
\end{eqnarray}
i.e. the magnetic moment grows linearly with the length $L=2^n$
obtained after $n$ RG steps
\begin{eqnarray}
\mu(L=2^n) && \opsimeq_{h \to 0} 2^n =L
\label{muLpurferro}
\end{eqnarray}
This corresponds as it should to the maximal magnetization per spin
$m=\frac{\mu(L)}{L}=1$.
As a consequence,
 the effective ferromagnetic coupling between two such magnetic moments
scales as
\begin{eqnarray}
J^{(n)}_{eff} \equiv J^{(n)} \mu(L=2^n) \mu(L=2^n) && \opsimeq_{h \to 0} L^{- (1+\sigma)} L^2  = L^{1-\sigma} = 2^{n (1-\sigma)}
\label{jmumuLpurferro}
\end{eqnarray}
in agreement with the classical Domain-Wall energy of Eq. \ref{edw}.

Deep in the ferromagnetic phase $h \to 0$,
the RG rule for the transverse field becomes (Eq. \ref{hrpur})
\begin{eqnarray}
h^R  \opsimeq_{h \to 0}   \frac{  h^2 }{J^{(0)} \mu^2  } = \frac{  h^2 }{J^{(0)}_{eff}  }
\label{hrpurferro}
\end{eqnarray}
After iteration over $n$ RG steps corresponding to the length $L=2^n$
\begin{eqnarray}
h(L=2^n)=h^{R^n}  \opsimeq_{h \to 0} 
  \frac{  h^{2^n} }{ J^{(n-1)}_{eff} (J^{(n-2)}_{eff})^2(J^{(n-3)}_{eff})^4...
(J^{(0)}_{eff})^{2^{n-1}}   }
=  \frac{  h^{2^n} }{ \prod_{k=0}^{n-1} (J^{(k)}_{eff})^{2^{n-1-k}}   }
\label{hrpurferroiter}
\end{eqnarray}
or equivalently in log-variables using $J^{(k)}_{eff}=2^{k (1-\sigma)}$ (Eq. \ref{jmumuLpurferro})
\begin{eqnarray}
\ln h(L=2^n) && \opsimeq_{h \to 0} 
 2^n \ln h - \sum_{k=0}^{n-1} 2^{n-1-k} \ln (J^{(k)}_{eff})
\nonumber \\
&& = 2^n \left[ \ln h - \sum_{k=0}^{n-1} 2^{-1-k} k (1-\sigma) \ln 2  \right]
\nonumber \\
&& = 2^n \left[ \ln h - (1-\sigma) \ln 2 \left( 1-(n+1)2^{-n} \right)  \right]
\nonumber \\
&& = L \left[ \ln h - (1-\sigma) \ln 2  \right]
+   (1-\sigma) (\ln L+ \ln 2)
\label{hrpurferrolog}
\end{eqnarray}
At leading order, the renormalized transverse field
is thus attracted exponentially in $L$ towards zero
\begin{eqnarray}
 h(L) && \opsimeq_{h \to 0} \left(\frac{h}{2^{1-\sigma}} \right)^L (2L)^{1-\sigma}
\label{hrpurferroleadingL}
\end{eqnarray}
so that it becomes smaller and smaller with respect to the effective ferromagnetic coupling of Eq. \ref{jmumuLpurferro}, i.e. the ferromagnetic fixed point $h=0$ is attractive.

\subsection{ RG rule for the control parameter $K$  }

It is clear that the important parameter of the model
is the ratio between the effective ferromagnetic coupling 
$J^{(0)} \mu^2$ and the transverse field $h$
\begin{eqnarray}
K \equiv \frac{J^{(0)} \mu^2 }{h}
\label{K0}
\end{eqnarray}
After one RG step, this control parameter is renormalized into
(Eqs \ref{hrpur} and \ref{RGmupur} )
\begin{eqnarray}
K^{R} && \equiv \frac{J^{(1)} (\mu^R)^2 }{h^{R}}
 =  2^{-1-\sigma} K \frac{ \left(  K + \sqrt{ K^2+4  } \right)^2}{ \sqrt{ K^2+4 }}
\equiv \phi_{\sigma}(K)
\label{KRpower}
\end{eqnarray}
The attractive paramagnetic fixed point corresponds to $K=0$,
whereas the attractive ferromagnetic fixed point corresponds to $K=+\infty$.
From now on, we focus on the unstable fixed point between them
 and on its critical properties.

\subsection{ Location of the critical point $K_c(\sigma)$}

The critical point $K_c$ corresponds to the non-trivial fixed point
$K_c=\phi_{\sigma}(K_c)$ of the RG rule of Eq. \ref{KRpower} leading to
\begin{eqnarray}
 (2^{\sigma}-K_c) \sqrt{ K_c^2+4 }= K_c^2+2  
\label{Kceq}
\end{eqnarray}
So keeping the condition $K_c<2^{\sigma} $,
 we may take the square to obtain the cubic equation
\begin{eqnarray}
K_c^3-2^{\sigma-1} K_c^2+4 K_c-2(2^{\sigma}-2^{-\sigma}) =0
\label{Kceqcubic}
\end{eqnarray}
In terms of the standard Cardano notations for cubic equations
\begin{eqnarray}
p &&  = 4 - \frac{4^{\sigma-1}}{3}
\nonumber \\
q && = 2^{1-\sigma}- \frac{2^{\sigma+2}}{3}- \frac{2^{3\sigma-2}}{27}
\nonumber \\
\sqrt{\frac{q^2}{4}+ \frac{p^3}{27}} && = (2^{-\sigma}+2^{\sigma-1}) \sqrt{1+\frac{4^{\sigma}}{27}}
\label{pq}
\end{eqnarray}
and
\begin{eqnarray}
u_{+} && = \left( -\frac{q}{2} + \sqrt{\frac{q^2}{4}+ \frac{p^3}{27}} \right)^{\frac{1}{3}}
\nonumber \\
&& =\left( - 2^{-\sigma}+ \frac{2^{\sigma+1}}{3}+ \frac{2^{3\sigma-3}}{27} 
+ (2^{-\sigma}+2^{\sigma-1}) \sqrt{1+\frac{4^{\sigma}}{27}}  \right)^{\frac{1}{3}}
\nonumber \\
u_{-} && = -{\rm sgn}(p)  \left\vert -\frac{q}{2} - \sqrt{\frac{q^2}{4}+ \frac{p^3}{27}} \right\vert^{\frac{1}{3}} \nonumber \\
&& = -{\rm sgn}(2+\frac{\ln 3}{2 \ln 2}-\sigma)  \left\vert
 - 2^{-\sigma}+ \frac{2^{\sigma+1}}{3}+ \frac{2^{3\sigma-3}}{27} 
- (2^{-\sigma}+2^{\sigma-1}) \sqrt{1+\frac{4^{\sigma}}{27}} \right\vert^{\frac{1}{3}}
\label{uv}
\end{eqnarray}
one obtains that the only real solution of Eq. \ref{Kceqcubic}
 reads
\begin{eqnarray}
K_c(\sigma)= \frac{2^{\sigma-1}}{3}+u_++u_-
\label{kcsigma}
\end{eqnarray}
It is a growing function of $\sigma$ with the values
\begin{eqnarray}
K_c(\sigma \to 0) && = \sigma \ln 2+O(\sigma^2)
 \nonumber \\
 K_c(\sigma=\frac{1}{4})&& =0.177439
 \nonumber \\
 K_c(\sigma=\frac{1}{2})&& =0.364946
\nonumber \\
K_c(\sigma=\frac{2}{3})&& =0.497043
\nonumber \\
K_c(\sigma=1)&& =0.783243
\label{kcsigmaspecial}
\end{eqnarray}

\subsection{ Correlation length exponent $\nu(\sigma)$ }

The correlation length exponent $\nu$ measures the instability
of the linearized RG flow around the critical point.
It is determined by the derivative of the RG flow of Eq. \ref{KRpower}
\begin{eqnarray}
2^{\frac{1}{\nu}} =
 \phi_{\sigma}'(K_c) =\frac{ 2  (2+K_c\sqrt{ K_c^2+4  }) }{( K_c^2+4)}
\label{KRpowerderic}
\end{eqnarray}

This yields that $\nu(\sigma)$ is a decaying function of $\sigma$ with the values
\begin{eqnarray}
\nu(\sigma \to 0) && \simeq \frac{1}{\sigma}+O(1)
 \nonumber \\
 \nu(\sigma=\frac{1}{4})&& =4.4406
 \nonumber \\
 \nu(\sigma=\frac{1}{2})&& =2.45131
\nonumber \\
\nu(\sigma=\frac{2}{3})&&=1.9602
\nonumber \\
\nu(\sigma=1)&&=1.482
\label{nusigmaspecial}
\end{eqnarray}

\subsection{ Magnetic exponent $x(\sigma)$ }

At criticality $K_c$, the RG rule for the renormalized 
 magnetic moment of Eq. \ref{RGmupur} 
\begin{eqnarray}
\mu_{R} 
&& =\mu \sqrt{2}   \sqrt{ 1+\frac{ K_c }{\sqrt{ K_c^2+4 } } }
\label{RGmupurcriti}
\end{eqnarray}
yields after $n$ RG steps corresponding to the length $L=2^n$
\begin{eqnarray}
\mu(L=2^n)  = \mu \left(\sqrt{2}   \sqrt{ 1+\frac{ K_c }{\sqrt{ K_c^2+4 } } }  \right)^n 
 =L^{1- x}
\label{mrpurcriti}
\end{eqnarray}
with the magnetic exponent 
\begin{eqnarray}
x(\sigma) && = 1- \frac{\ln \left(\sqrt{2}   \sqrt{ 1+\frac{ K_c }{\sqrt{ K_c^2+4 } } }  \right)}{\ln 2}
\nonumber \\
&& = \frac{1-\sigma}{4}+ \frac{\ln(K_c^2+4)}{8 \ln 2}
\label{xpur}
\end{eqnarray}
It is a decaying function of $\sigma$ with the values
\begin{eqnarray}
x(\sigma \to 0) && = \frac{1}{2}- \frac{\sigma}{4}+O(\sigma^2)
 \nonumber \\
 x(\sigma=\frac{1}{4})&& =0.438914
 \nonumber \\
 x(\sigma=\frac{1}{2})&& =0.380907
\nonumber \\
x(\sigma=\frac{2}{3})&& =0.344141
\nonumber \\
x(\sigma=1)&& =0.275732
\label{betanusigmaspecial}
\end{eqnarray}

\subsection{ Dynamical exponent $z(\sigma)$ }

At criticality $K_c$, the RG rule for the transverse field
$h$ of Eq. \ref{hrpur} 
\begin{eqnarray}
h^R  = h  \frac{ 2  }
{ K_c + \sqrt{ K_c^2+4  } }
\label{hrpurcriti}
\end{eqnarray}
yields after $n$ RG steps corresponding to the length $L=2^n$
\begin{eqnarray}
h(L=2^n)  = h \left( \frac{ 2  }
{ K_c + \sqrt{ K_c^2+4  } } \right)^n  = h L^{-z}
\label{hrpurcritisol}
\end{eqnarray}
with the dynamical exponent 
\begin{eqnarray}
z(\sigma) && =
 \frac{ \ln \left(\frac{ K_c + \sqrt{ K_c^2+4  } }{ 2  }\right) }{\ln 2}
= \frac{\sigma-1}{2} + \frac{\ln(K_c^2+4)}{4 \ln 2}
\label{zpur}
\end{eqnarray}

By construction at criticality, the dynamical exponent $z$ also describes
the scaling of the effective ferromagnetic coupling
\begin{eqnarray}
L^{-z} = J_{eff}(L) = L^{-1-\sigma} \mu^2(L) = L^{1-\sigma-2 x} 
\label{zjferro}
\end{eqnarray}
i.e. it is directly related to the magnetic exponent $x$ of Eq. \ref{xpur}
\begin{eqnarray}
 z(\sigma) = \sigma-1+2x(\sigma)
\label{zx}
\end{eqnarray}
with the values
\begin{eqnarray}
z(\sigma \to 0) && = \frac{\sigma}{2}+O(\sigma^2)
 \nonumber \\
 z(\sigma=\frac{1}{4})&& =0.127828
 \nonumber \\
 z(\sigma=\frac{1}{2})&& =0.261814
 \nonumber \\
z(\sigma=\frac{2}{3})&&=0.354949
\nonumber \\
z(\sigma=1)&&=0.551463
\label{zsigmaspecial}
\end{eqnarray}

\subsection{  Correlation exponent $\eta(\sigma)$ }

  Within the real-space RG procedure, the two-point spatial correlation function
  scales as the square of the magnetization at the scale where the two
spins are merged into a single renormalized spin
  \begin{eqnarray}
  C(L=2^n) \equiv <\sigma^x_1 \sigma^x_L> 
 \propto \left(\frac{\mu(L)}{L}  \right)^2 \propto L^{-2x}
  \label{correpurcriti}
  \end{eqnarray}
As a consequence, the exponent $\eta$ of the 
correlation function defined by
  \begin{eqnarray}
C(r) \propto r^{-(d-2+z+\eta})
  \label{defeta}
  \end{eqnarray}
after taking into account Eq. \ref{zx}, takes the simple explicit value
  \begin{eqnarray}
\eta(\sigma)=2-\sigma
  \label{etax}
  \end{eqnarray}
that coincides with the mean-field value of Eq. \ref{etamf}.
The fact that $\eta$ keeps its mean-field value $(2-\sigma)$ 
even in the non-mean-field region until it reaches its Short-Ranged value $\eta_{SR}$
is expected for the Long-Ranged quantum Ising model \cite{dutta2001}
 for the same reasons as in the classical case \cite{luijtenblote2002}.

\subsection{ Ground state energy }

In the pure case, the renormalization rule of Eq. \ref{rgegs}
for the contribution of a single block to the ground state energy simplifies into
\begin{eqnarray}
 e_{R} 
&& =- \frac{ J^{(0)}\mu^2+ \sqrt{ (J^{(0)}\mu^2 )^2+4h^2 }  }{2}
\label{rgegspur}
\end{eqnarray}

For a chain of $L=2^n$ spins, the ground state energy can be obtained by
summing all the contributions of the successive projections for the blocks 
up to the last step where the single remaining spin has for contribution
$(-h_{R^n} )$
\begin{eqnarray}
 E^{GS}(L=2^n,K) && = - \frac{L}{2} e_{R} - \frac{L}{4} e_{R^2}- \frac{L}{8} e_{R^3}-...
- e_{R^{n-1}}-h_{R^n}
\nonumber \\ &&
= -  \sum_{k=0}^{n-1} \frac{L}{2^{1+k}} e_{R^k} -h_{R^n}
\label{egspure}
\end{eqnarray}
In the paramagnetic limit $J=0=K$, where the transverse fields are not renormalized
(Eq. \ref{hrpurpara}), one recovers as it should the contribution $(-h)$ for each independent spin
\begin{eqnarray}
 E^{GS}_{para}(L=2^n,J=0)&&
= -  \sum_{k=0}^{n-1} \frac{L}{2^{1+k}} h -h = -h L
\label{egspurepara}
\end{eqnarray}
In the ferromagnetic limit $h=0$,$K=+\infty$, 
where the effective ferromagnetic coupling between two such magnetic moments
scales as Eq \ref{jmumuLpurferro},
one obtains 
\begin{eqnarray}
 E^{GS}_{ferro}(L=2^n,h=0)&&
= -  \sum_{k=0}^{n-1} \frac{L}{2^{1+k}}   2^{k (1-\sigma)}
= -  \frac{1}{2 (1-2^{-\sigma})} (L-L^{1-\sigma})
\label{egspureferro}
\end{eqnarray}
as it should : the coefficient of the extensive term corresponds to
 the half of the sum of the couplings starting from one site, for instance the first one,
i.e. $(\sum_{i=2}^{+\infty} J_{1,i})/2 $, whereas the correction to extensivity in $L^{1-\sigma}$
represents the 'missing' couplings of the region $j>L$.

At criticality $K=K_c$, where the renormalized transverse-fields involve the dynamical exponent $z$ (Eq. \ref{hrpurcritisol} and Eq. \ref{zpur}), the ground state energy reads
\begin{eqnarray}
 E^{GS}_{criti}(L=2^n,K_c) && 
= -  \sum_{k=0}^{n-1} \frac{L}{2^{1+k}} h_{R^k} \frac{ K_{R^k}+ \sqrt{ (K_{R^k})^2+4 }  }{2} -h_{R^n}
\nonumber \\ &&
=  -  \sum_{k=0}^{n-1} \frac{L}{2^{1+k}} h 2^{-z k} \frac{ K_c+ \sqrt{ K_c^2+4 }  }{2}
 - h 2^{-z n}
\nonumber \\ &&
= - h \sum_{k=0}^{n-1} \frac{L}{2^{1+k}}  2^{-z k} 2^{z} - h 2^{-z n}
\nonumber \\ &&
= - h \frac{2^{2 z}}{2^{1+z}-1} L+   \frac{(2^{ z}-1)^2}{2^{1+z}-1} L^{-z}
\label{egspurecriti}
\end{eqnarray}
i.e. the finite-size correction to the ground state energy per spin
\begin{eqnarray}
\frac{ E^{GS}_{criti}(L=2^n,K_c)}{L} && 
= - h \frac{2^{2 z}}{2^{1+z}-1} +   \frac{(2^{ z}-1)^2}{2^{1+z}-1} L^{-1-z}
\label{egspurecritiint}
\end{eqnarray}
is of order $  L^{-1-z}$ as expected.

\subsection{ Discussion }

In summary, the real-space RG
 procedure for the pure Dyson hierarchical quantum Ising model
has the advantage of being explicitly solvable for the various observables
and to yield reasonable approximated values for the critical exponents. 
Besides the simple anomalous dimension $\eta=2-\sigma$
 (Eq. \ref{etax}) reproduced both in the mean-field region and in the non-mean-field region as it should, it is interesting to consider
 some specific values of $\sigma$ to compare with
previous approaches :

(i) for the non-mean-field value $\sigma=1$,
the dynamical exponent $z(\sigma=1) \simeq 0.55 $,
and the anomalous dimension $\eta(\sigma)=1$ are close
to the Monte-Carlo measure quoted in Eq. \ref{zpureLRnume} of the Introduction,
even if the correlation length $\nu(\sigma=1)=1.48$ is less good
with respect to Eq. \ref{zpureLRnume}.

(ii) at the upper critical value $\sigma_u=2/3$, 
the dynamical exponent $z(\sigma=\frac{2}{3}) \simeq 0.35 $
and the magnetic exponent $x(\sigma=\frac{2}{3}) \simeq 0.34 $
are close to the expected values of Eqs \ref{sigmaunuz} and \ref{sigmaux},
even if the value for the correlation length exponent $\nu(\sigma=\frac{2}{3}) \simeq
1.96 $ is again not so good with respect to Eq. \ref{sigmaunuz}.

(iii) at first order in the expansion near $\sigma \to 0$, 
 the critical exponents $\nu(\sigma)$, $x(\sigma)$, $z(\sigma)$
and the critical point location $K_c(\sigma)$ 
found above by the real-space RG procedure actually
coincide with the mean-field values recalled in Appendix \ref{sec_meanfield}.
Further work is needed to understand if the renormalization procedure
can be modified to describe correctly the whole mean-field region
 $0<\sigma<\sigma_u=2/3$ and in particular its anomalous finite-size scaling properties.
Indeed, the anomalous finite-size scaling properties of the mean-field region
of classical statistical physics models
 has a long history that has been re-interpreted recently
(see \cite{bb2012,bb2013,bb2014,bb2015} and references therein), and it would be 
very interesting to see how it can emerge explicitly within some real-space RG.
For the present case, the mean-field critical properties 
can actually be reproduced by the real-space RG flow for the control parameter $K=\frac{J^{(0)} \mu^2 }{h}$
(instead of Eq. \ref{KRpower})
\begin{eqnarray}
K^{R}_{MF} && = 2^{-\sigma} \frac{K}{1-K}
\label{rgKmf}
\end{eqnarray}
with the critical point location $K_c(\sigma)=1- 2^{-\sigma} $ and the correlation length
exponent $\nu(\sigma)=\frac{1}{\sigma} $.
More precisely, the corresponding rules for the transverse field and the magnetic moment read (instead of Eqs \ref{hrpur} and \ref{RGmupur})
\begin{eqnarray}
h^{R}_{MF} && = h \sqrt{1-K}
\nonumber \\
\mu^{R}_{MF} && = \mu \sqrt{\frac{2}{\sqrt{1-K}}}
\label{rghmumf}
\end{eqnarray}
corresponding to the dynamical exponent $z(\sigma)=\frac{\sigma}{2}$
and to the magnetic exponent $x(\sigma)=\frac{1}{2}-\frac{\sigma}{4} $.
Within the real-space RG perspective, it is not clear to us
what arguments should be used to justify that this mean-field RG flow has to be preferred 
to describe the critical point in the whole region $0 <\sigma\leq \sigma_u=2/3$.

\section{ Dyson hierarchical chain with random transverse fields }

\label{sec_random}

In this section, we consider the Dyson hierarchical chain of Eq. \ref{recDyson}
where the transverse fields $h_i$ are independent random variables
drawn with the flat distribution over the interval $[0,W]$
\begin{eqnarray}
P(h) = \frac{ \theta(0 < h \leq W) }{W}
\label{flat}
\end{eqnarray}
which is the distribution previously used in the Strong Disorder RG study \cite{strongLR}.
The parameter $W$ which represents the typical scale of the initial transverse fields
 is thus the control parameter of the transition.
It will be actually more convenient to use
\begin{eqnarray}
\theta \equiv \ln W
\label{theta}
\end{eqnarray}
as in the previous Strong Disorder RG study \cite{strongLR}.

\subsection{ RG rules deep in the ferromagnetic phase $ W \to 0$}

When the transverse fields $(h_{2i-1},h_{2i})$ are small 
with respect to the ferromagnetic term $J^{(0)} \mu_{2i-1} \mu_{2i}$,
the RG rule for the magnetic moment (Eq. \ref{RGmu}) is simply additive
\begin{eqnarray}
\mu_{R(2i)} && =  \mu_{2i-1}+ \mu_{2i}
\label{RGmuferrorandom}
\end{eqnarray}
As a consequence, after $n$ RG steps corresponding to the length $L=2^n$,
the magnetic moment is given by the number of spins
\begin{eqnarray}
\mu(L) && = L
\label{RGmuferropr}
\end{eqnarray}
and the effective ferromagnetic coupling between two such magnetic moments
scales as
\begin{eqnarray}
J_{eff}^{(n)} \equiv J^{(n)} \mu(L=2^n) \mu(L=2^n) \simeq  2^{n(1-\sigma)} = L^{1-\sigma}
\label{jmumuLpurferror}
\end{eqnarray}
in agreement with the classical Domain-Wall energy of Eq. \ref{edw}.

The renormalization rule for the transverse field (Eq. \ref{rgh}) becomes
\begin{eqnarray}
 h_{R(2i)} \simeq \frac{  h_{2i-1} h_{2i} }{J^{(0)} \mu_{2i-1} \mu_{2i} }
= \frac{  h_{2i-1} h_{2i} }{J^{(0)}_{eff}  }
\label{rghferro}
\end{eqnarray}
Upon iteration in log-variables, one can follow
the same steps as in the pure case (Eqs \ref{hrpurferroiter} and \ref{hrpurferrolog})
and one obtains
\begin{eqnarray}
\ln h(L) && \simeq \sum_{i=1}^L \ln h_i - L (1-\sigma) \ln 2 
+   (1-\sigma) (\ln L+ \ln 2)
\label{hrrandomferrolog}
\end{eqnarray}

For large $L$, the Central Limit theorem yields 
that the typical value 
\begin{eqnarray}
\ln h_{typ}(L) && \equiv \overline{ \ln h(L) } 
\nonumber \\
&& \simeq L \left( \overline{ \ln h_i } -  (1-\sigma) \ln 2 \right)
+   (1-\sigma) (\ln L+ \ln 2)
\label{hrrandomferrologav}
\end{eqnarray}
 becomes smaller and smaller with respect to the effective ferromagnetic coupling of Eq. \ref{jmumuLpurferror}, i.e. the ferromagnetic fixed point is attractive.
In this ferromagnetic phase, the variance grows linearly (Eq. \ref{hrrandomferrolog})
\begin{eqnarray}
Var[\ln h(L)] = L Var[ \ln h_{i} ]
\label{rghferrologvar}
\end{eqnarray}

\subsection{ RG rules deep in the paramagnetic phase $ W \to +\infty$  }

\label{sec_deeppara}

When the ferromagnetic term $J^{(0)} \mu_{2i-1} \mu_{2i}$
is small with respect to the transverse fields $(h_{2i-1},h_{2i})$
the RG rule for the transverse field (Eq. \ref{rgh}) becomes
\begin{eqnarray}
 h_{R(2i)} \simeq \frac{ 2 h_{2i-1} h_{2i} }{
h_{2i-1}+h_{2i} + \vert h_{2i-1}-h_{2i} \vert } = {\rm min} (h_{2i-1},h_{2i} )
\label{rghdespara}
\end{eqnarray}
and the RG rule for the magnetic moment (Eq. \ref{RGmu}) reads
\begin{eqnarray}
\mu_{R(2i)} 
&& = \sqrt{ \frac{1- sgn(h_{2i-1}-h_{2i}) }{2}}   \mu_{2i-1}
+ \sqrt{ \frac{1+ sgn(h_{2i-1}-h_{2i}) }{2}}  \mu_{2i}
\nonumber \\
&& = \theta(h_{2i-1}<h_{2i} ) \mu_{2i-1}+ \theta(h_{2i-1}>h_{2i} ) \mu_{2i}
\label{RGmudespara}
\end{eqnarray}
In this paramagnetic limit, the renormalized spin is thus simply
the spin with the smallest transverse field that survives with its properties,
 whereas the spin with the highest transverse field is decimated and disappears.

So after $n$ RG steps corresponding to the length $L=2^n$,
the magnetic moment has simply kept its initial value
\begin{eqnarray}
\mu(L=2^n) && = 1
\label{RGmupara}
\end{eqnarray}
so that the effective magnetic coupling also keeps its bare value
\begin{eqnarray}
 J_{eff}(L) = J^{(p)} = 2^{-(1+\sigma)p}=L^{-1-\sigma}
\label{jeffparar}
\end{eqnarray}

The renormalized transverse field is
given by the minimum value among the $L=2^n$ initial random variables
\begin{eqnarray}
 h(L=2^n) =  {\rm min} (h_{1},h_{2},..,h_{L} ) 
\label{rghparamin}
\end{eqnarray}
Its probability distribution $P_L(h)$ reads thus in terms of the initial distribution $P_1(h)$ of Eq. \ref{flat}
\begin{eqnarray}
 P_L(h) && = L P_1(h) \left[ \int_h^{+\infty} dx P_1(x)  \right]^{L-1}
\nonumber \\
&& = L \frac{ \theta(0 \leq h \leq W) }{W}
 \left[ \frac{ W -h }{W}  \right]^{L-1}
\label{phpara}
\end{eqnarray}
For large $L$, it converges towards the exponential distribution
(as a particular case of the Fr\'echet distribution)
\begin{eqnarray}
 P_L(h) && \opsimeq_{L \to +\infty}   \frac{ L }{W} e^{-  \frac{L}{W} h }
\label{phparafrechet}
\end{eqnarray}
The characteristic scale
\begin{eqnarray}
h_{typ}(L)= \frac{W}{L}
\label{rghparaminboxtyp}
\end{eqnarray}
decays, but remains bigger that the effective ferromagnetic coupling of Eq. \ref{rghparamin}, so that the paramagnetic fixed point is attractive.

The approximate renormalization rules derived here in the limit $W \to +\infty$ 
 are similar to the Strong Disorder RG rules \cite{strongLR}.
However the numerical study described below yields that the paramagnetic phase
near the critical point is described by other scaling behavior.

\subsection{ Numerical study of the Critical Point  }

To study the critical properties, we have used two methods :

(i) on one hand, we have generated $n_s=4.10^3$ disordered samples
of the Dyson chain containing $N=2^{24} \simeq 1.68 \times 10^7$ spins,
corresponding to $n=24$ generations,
and we have applied numerically the RG rules of Eqs \ref{rgh} and \ref{RGmu}.

(ii) on the other hand, we have used the so-called standard 'pool method' :
the idea is to represent the joint probability distribution
$P_n(h,\mu)$ of the renormalized transverse field $h$ and of
the magnetic moment $\mu$ of a renormalized quantum spin at generation $n$
 by a pool of $M$ realizations $(h_n^{(i)},\mu_n^{(i)})$ where $i=1,2,...,M$.
The pool at generation $(n+1)$ is then constructed as follows :
each new realization $(h_{n+1}^{(i)},\mu_{n+1}^{(i)})$ is obtained by choosing 
$2$ ancestors spins at random from the pool of generation $n$ and by applying
the renormalization rules of Eqs \ref{rgh} and \ref{RGmu}.
The numerical results presented below have been obtained with a pool of size
$M=3.10^7$ iterated up to $n=100$ generations.
The pool method allows to study much bigger sizes and statistics,
but introduces some truncation related to the size of the pool
in the probability distributions.
We have checked that in the critical region, the results of (ii) are in agreement
with (i) for the common sizes.

In both methods, for each renormalization step corresponding to the lengths $L=2^n$,
we have analyzed the statistical properties of the renormalized transverse fields 
$h_L$ and of the renormalized magnetic moments $\mu_L$. 
More precisely, we have measured the RG flows of the corresponding typical values 
\begin{eqnarray}
\ln h_L^{typ} && \equiv \overline{ \ln h_L } 
\nonumber \\
\ln \mu_L^{typ} && \equiv \overline{ \ln \mu_L }  
\label{deftyp}
\end{eqnarray}
as a function of the length $L$ for various values of the control parameter $\theta$
 of the initial distribution of Eq. \ref{flat}.
We have studied the critical properties for the two values $\sigma=2/3$ and $\sigma=1$
with the following critical point
\begin{eqnarray}
\theta_c \left( \sigma=\frac{2}{3} \right) && \simeq 2.80725
\nonumber \\
\theta_c \left( \sigma=1 \right) && \simeq 2.14875
\label{thetac}
\end{eqnarray}

\subsection{ RG flows of the magnetic moment $\mu_L$  }

\begin{figure}[htbp]
 \includegraphics[height=6cm]{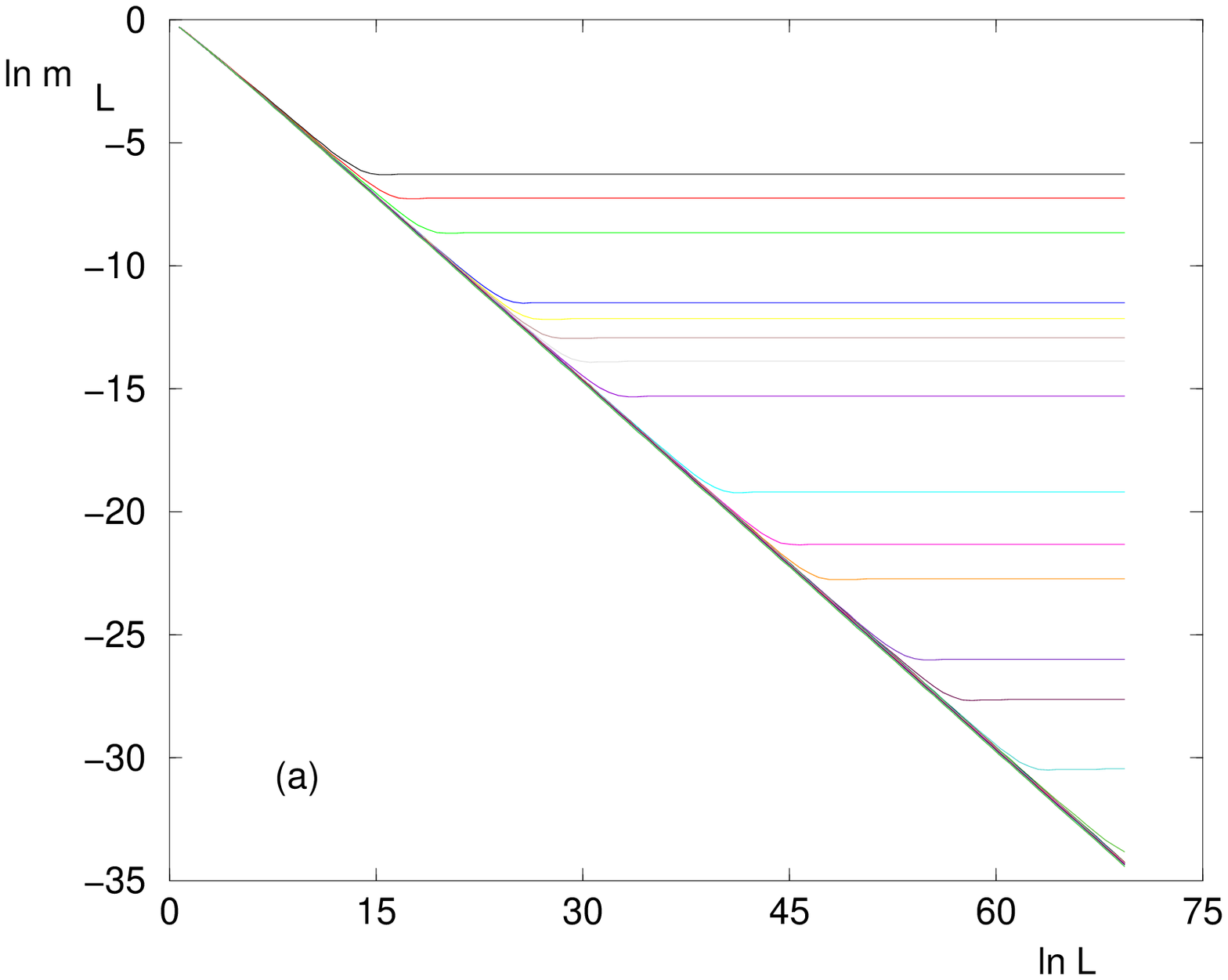}
\hspace{1cm}
 \includegraphics[height=6cm]{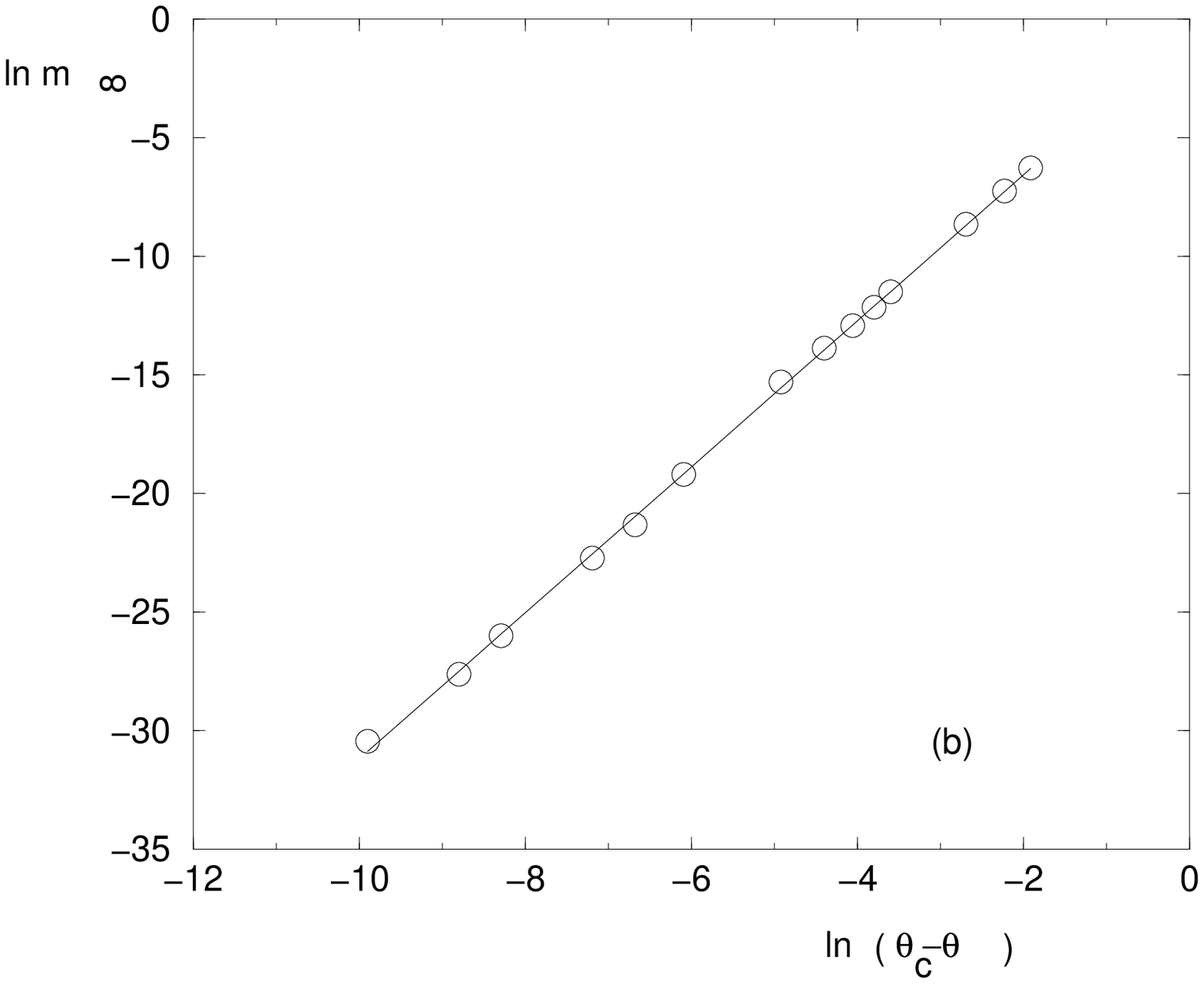}
\caption{ Case $\sigma=\frac{2}{3}$  \\
 (a) RG flow of the magnetization $m_L^{typ} \equiv \frac{\mu_L^{typ}}{L}$
in log-log plot for various control parameter $\theta$ \\ 
(i) in the ferromagnetic phase $\theta<\theta_c$, the flow converges towards
a finite magnetization $m_{\infty}(\theta)$ ; \\
(ii) in the paramagnetic phase $\theta>\theta_c$, the slope is $(-1/2)$ ; \\
(iii) at criticality, the slope $(-x(\sigma=2/3)) \simeq -0.5$ is not distinguishable
from the paramagnetic slope.  \\
(b)  $ \ln (m_{\infty}(\theta)) $ as a function of $\ln (\theta_c-\theta)$ :
 the slope corresponds to the exponent $\beta(\sigma=\frac{2}{3}) \simeq 3 $  }
\label{figflowm}
\end{figure}

On Fig. \ref{figflowm} (a), we show the RG flows of 
the magnetization $m_L^{typ} \equiv \frac{\mu_L^{typ}}{L}$  (Eq. \ref{deftyp})
as a function of the length $L$ for various control parameter $\theta$.
They can be summarized as :
\begin{eqnarray}
m_L^{typ} \vert_{\theta<\theta_c} && \opsimeq_{L \to +\infty}   m_{\infty}(\theta) \nonumber \\
 m_L^{typ} \vert_{\theta=\theta_c}  && \oppropto_{L \to +\infty}  L^{-x(\sigma)}    \nonumber \\
 m_L^{typ} \vert_{\theta>\theta_c} && \oppropto_{L \to +\infty} A(\theta) L^{-\frac{1}{2}} 
\label{mutypflow}
\end{eqnarray}

On the ferromagnetic side, we obtain that the asymptotic finite magnetization 
vanishes as a power-law (see Fig. \ref{figflowm} (b)) 
\begin{eqnarray}
m_{\infty}(\theta)  \oppropto_{\theta \to \theta_c^-} (\theta_c-\theta)^{\beta(\sigma)} 
\label{defbeta}
\end{eqnarray}
and we measure the same value for $\sigma=2/3$ and $\sigma=1$
\begin{eqnarray}
\beta(\sigma=\frac{2}{3}) \simeq 3 \simeq \beta(\sigma=1)
\label{resbeta}
\end{eqnarray}

On the paramagnetic side, it is not clear that the amplitude $A(\theta)$ diverges,
and correspondingly at criticality, the magnetic exponent $x(\sigma)$
cannot be distinguished from the paramagnetic phase value $1/2$
\begin{eqnarray}
x(\sigma=\frac{2}{3}) \simeq 0.5 \simeq x(\sigma=1)
\label{resx}
\end{eqnarray}
As a consequence, the finite-size scaling matching between Eq. \ref{resbeta}
and \ref{resx} corresponds to the finite-size correlation length exponent
$\nu_{FS}=\frac{\beta}{x} $
\begin{eqnarray}
\nu_{FS}(\sigma=\frac{2}{3}) \simeq 6 \simeq \nu_{FS}(\sigma=1)
\label{resnufss}
\end{eqnarray}

\subsection{ RG flow of the effective coupling $J_L$ }

In the present real-space RG framework, the effective coupling  $J_L$
is just the 'slave' of the magnetic moments,
and the typical value is given by
\begin{eqnarray}
 J_L^{typ} = L^{-1-\sigma} (\mu_{L}^{typ})^2 = L^{1-\sigma} (m_{L}^{typ})^2
\label{jslave}
\end{eqnarray}
The translation of Eq. \ref{mutypflow}
corresponds to the behaviors
\begin{eqnarray}
 J_L^{typ} \vert_{\theta<\theta_c} && \oppropto_{L \to +\infty} L^{1-\sigma}  m_{\infty}^2(\theta)
  \nonumber \\
 J_L^{typ} \vert_{\theta=\theta_c}  && \oppropto_{L \to +\infty} L^{1-\sigma-2x}= L^{-z(\sigma)}  \nonumber \\
 J_L^{typ} \vert_{\theta>\theta_c} && \oppropto_{L \to +\infty} L^{-\sigma} A^2(\theta)
\label{jtypflow}
\end{eqnarray}
Eq \ref{resx} yields that the critical dynamical exponent is given by
\begin{eqnarray}
z(\sigma)= \sigma-1+2x \simeq \sigma
\label{reszsigma}
\end{eqnarray}

\subsection{ RG flow of the renormalized transverse fields }

\begin{figure}[htbp]
 \includegraphics[height=6cm]{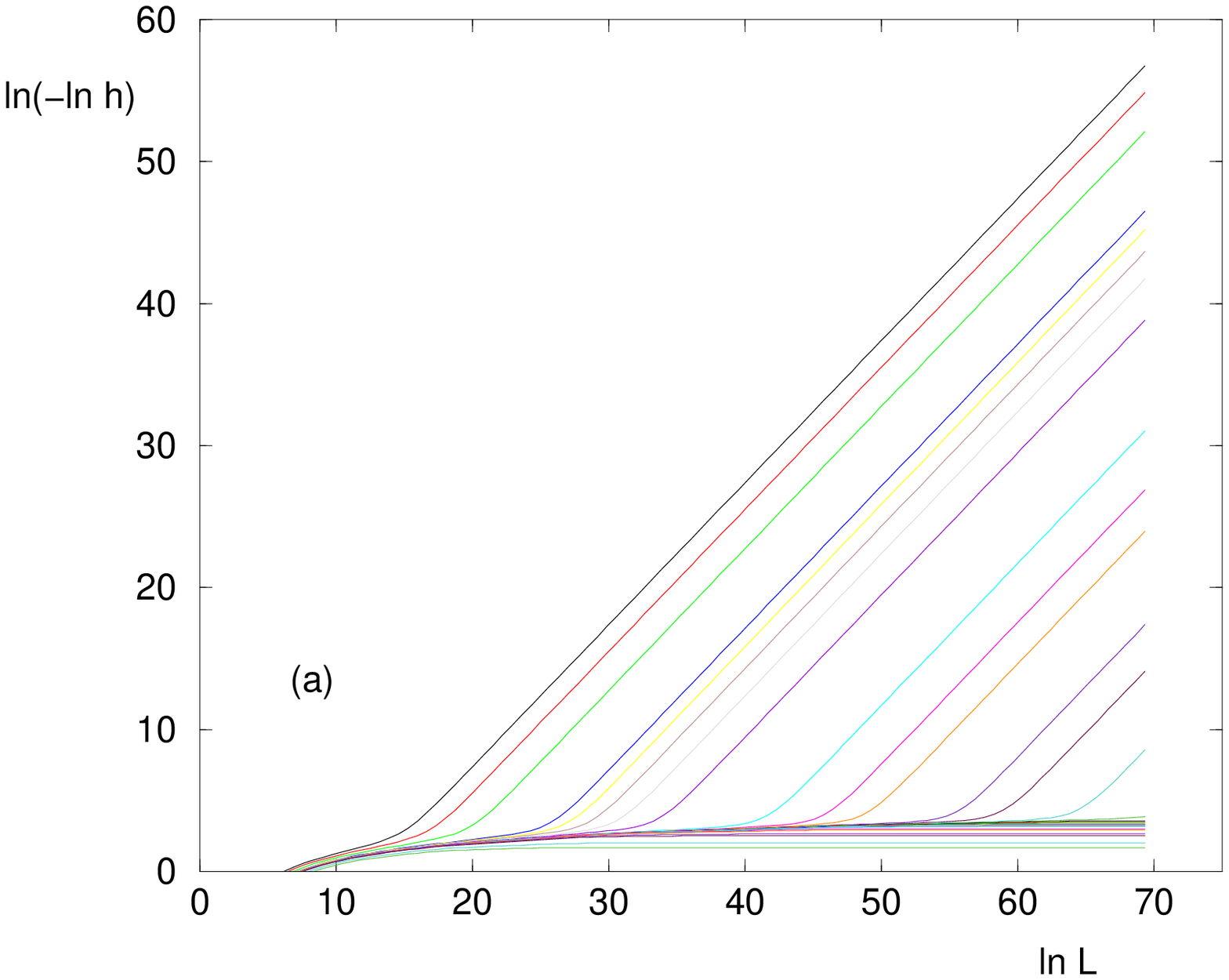}
\hspace{1cm}
 \includegraphics[height=6cm]{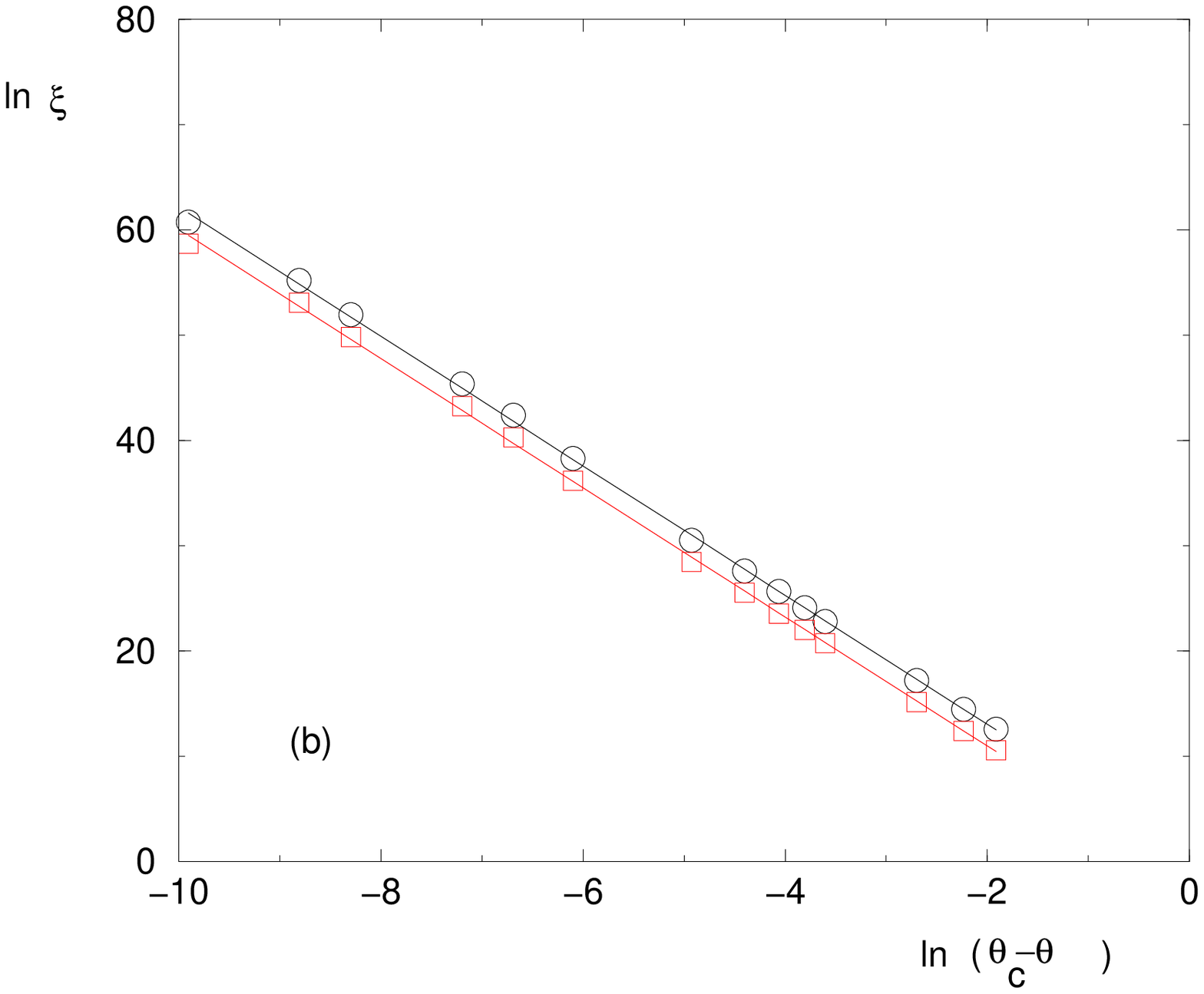}
\caption{ RG flow of the transverse fields for $\sigma=\frac{2}{3}$ \\
(a)  $\ln(- \ln h_L^{typ})$  as a function of $\ln L$ : \\
 in the ferromagnetic phase $\theta<\theta_c$, the slope unity 
allows to measure the correlation length $\xi_{h}(\theta)$ of Eq. \ref{htypflow}.  \\
(b)   $ \ln \xi_h(\theta)$ and $ \ln \xi_{width}(\theta)$ as a function of $\ln (\theta_c-\theta)$ :
 the slope corresponds to the value $\nu(\sigma=\frac{2}{3}) \simeq 6 $.
  }
\label{figflowh}
\end{figure}

At criticality, the typical renormalized transverse field 
involves the same dynamical exponent
as in Eq. \ref{reszsigma}
\begin{eqnarray}
 h_L^{typ} \vert_{\theta=\theta_c}  && \oppropto_{L \to +\infty} L^{- z(\sigma)} 
\label{htypflowcriti}
\end{eqnarray}

In the ferromagnetic phase, the exponential decay of the 
typical transverse field allows to define some correlation length $\xi_h(\theta)$
(see Fig. \ref{figflowh} (a))
\begin{eqnarray}
 \ln h_L^{typ} \vert_{\theta<\theta_c}  && \oppropto_{L \to +\infty} - \frac{L}{\xi_h(\theta)} 
\label{htypflow}
\end{eqnarray}
 The growth of the width of the probability distribution of $\ln h$
(Eq. \ref{rghferrologvar})
\begin{eqnarray}
\Delta_{\ln h_L} && \equiv  \sqrt{ \overline{ (\ln h_L)^2 } - (\overline{ \ln h_L })^2 }
\propto \left( \frac{L}{\xi_{width}} \right)^{\frac{1}{2}}
\label{defwidth}
\end{eqnarray}
also allows to define another correlation length $\xi_{width} $.
Their divergences
near criticality involve the same exponent $\nu(\sigma)$
(see Fig. \ref{figflowh} (b))
\begin{eqnarray}
\xi_h(\theta) && \oppropto_{\theta \to \theta_c} (\theta_c-\theta)^{- \nu(\sigma)} 
\nonumber \\
\xi_{width}(\theta) && \oppropto_{\theta \to \theta_c} (\theta_c-\theta)^{- \nu(\sigma)} 
\label{nuh}
\end{eqnarray}
Our numerical results for $\sigma=2/3$ and $\sigma=1$ point towards the same value
\begin{eqnarray}
\nu(\sigma=\frac{2}{3}) \simeq 6 \simeq \nu(\sigma=1)
\label{resnu}
\end{eqnarray}
in agreement with the finite-size scaling exponent of Eq. \ref{resnufss}.

\subsection{ Comparison with the previous Strong Disorder RG results
\cite{strongLR} }

Let us compare the results described above with the Strong Disorder RG study
 \cite{strongLR} :

(i) here the dynamical exponent is found to be $z(\sigma) \simeq \sigma$
instead of the value $z(\sigma) \simeq 1+\sigma$ obtained via Strong Disorder RG 
\cite{strongLR}. 
As a consequence of the relation of Eq. \ref{zx},
this difference of unity in $z$
is directly related to the difference between the magnetic exponent
$x(\sigma) \simeq 0.5$ obtained here
and $x =1$ found in Ref \cite{strongLR} (where more precisely the magnetization grows as
$m(L) \propto \frac{(\ln L)^2}{L}$). At the level of the renormalization rules,
the origin of this difference is the following :
whereas the present renormalization rules become equivalent to the Strong Disorder RG rules
in the limit $W \to +\infty$ (see section \ref{sec_deeppara}), 
they are not equivalent anymore
in the critical region. Indeed in
 the present framework, the magnetization is already of order $m(L) \propto L^{-1/2}$
in the paramagnetic phase, which can be seen as the result of Central Limit fluctuations.
 Since the magnetization at criticality cannot be smaller,
one obtains the bound $x \leq 1/2$.

(ii) instead of the essential singularity of Eq. \ref{xirandomSD} \cite{strongLR},
we obtain here conventional power-law scaling laws with
  finite but rather large correlation length exponent $\nu \simeq 6$,
and magnetic exponent $\beta \simeq 3$.

In summary, the Strong Disorder RG results obtained in \cite{strongLR}
for the Long-Ranged chain are based on the approximation that the quantum
fluctuations are negligible with respect to disorder fluctuations,
whereas the present block renormalization for the Dyson version
yields that quantum fluctuations
are important in the critical region.
Further work is needed to better understand whether
the differences in the results are entirely due to the different approximations
made in the renormalization rules, or whether the Dyson hierarchical version 
actually changes significantly the Long-ranged model and reduces the importance of rare events.

\section{Conclusion} 

\label{sec_conclusion}

In this paper, we have proposed to study via real-space renormalization
the Dyson hierarchical version of the quantum Ising chain with Long-Ranged power-law ferromagnetic couplings $J(r) \propto r^{-1-\sigma}$ and pure or random transverse fields. For the pure case, the RG rules have been explicitly solved as a function of the parameter $\sigma$, and we have compared the critical exponents with previous results of other approaches.
 For the random case, we have studied numerically the RG rules and compared the critical properties with the previous Strong Disorder Renormalization approach \cite{strongLR}.

Our conclusion is that the Dyson hierarchical idea initially 
developed for classical spins is also very useful for quantum spin models, 
in order to derive real-space renormalization rules. In the future, we hope to apply
this method to other quantum models.

\appendix

\section{ Reminder on the Mean-Field theory for the pure Long-Ranged model }

\label{sec_meanfield}

\subsection{ Thermodynamic exponents }

For the pure quantum Ising chain of Eq. \ref{hqising}, the mean-field theory
amounts to look for the uniform magnetization $m$
\begin{eqnarray}
< \sigma^x_i > && = m
\nonumber \\
< \sigma^z_i > && = \sqrt{1-m^2} 
\label{mfm}
\end{eqnarray}
that minimizes the ground state energy per spin
\begin{eqnarray}
e_{GS}(m) = \frac{E_{GS}(m)}{N} = -   h \sqrt{1-m^2}   - \frac{1}{2} J_{tot} m^2
\label{egsmf}
\end{eqnarray}
where $J_{tot}$ represents the sum of all couplings linked to a given spin
\begin{eqnarray}
 J_{tot} =\sum_{j \ne 1} J_{1,j}
\label{jtot}
\end{eqnarray}
For instance for the Dyson model of Eq. \ref{powerk}, it is given by
\begin{eqnarray}
 J_{tot}^{Dyson} =\sum_{k=0}^{+\infty} 2^k J^{(k)} = \sum_{k=0}^{+\infty} 2^{-\sigma k } = \frac{1}{1-2^{-\sigma}}
\label{jtotdyson}
\end{eqnarray}
The minimization of Eq. \ref{egsmf}
\begin{eqnarray}
0=\partial_m e_{GS}(m) =    h \frac{m}{\sqrt{1-m^2}}   -  J_{tot} m
\label{egsmfderi}
\end{eqnarray}
yields that the critical transverse field is
\begin{eqnarray}
h_c=J_{tot}
\label{hcmf}
\end{eqnarray}
and that the optimal magnetization reads
\begin{eqnarray}
m && =0 \ \ \ \ \ \ \ \ \ \ \ \ \ \ \ \ \ \ \ \ \  {\rm for } \ \ h>h_c
\nonumber \\
m && =\sqrt{1- \left(\frac{h}{h_c}\right)^2} \ \ \ {\rm for } \ \ h<h_c
\label{msol}
\end{eqnarray}
corresponding to the usual mean-field thermodynamic exponents
\begin{eqnarray}
\beta_{MF}=\frac{1}{2} 
\nonumber \\
 \alpha_{MF}=0 
\label{mfthermo}
\end{eqnarray}
independently of the short-range or long-ranged nature of the couplings.

\subsection{ Correlation exponents for the short-ranged case  }

As recalled in the Introduction,
the short-ranged model is characterized by the dynamical exponent
(both in mean-field and outside mean-field) \cite{sachdev}
\begin{eqnarray}
z^{SR}=1
\label{zsr}
\end{eqnarray}

The correlation length exponent $\nu$ and the exponent $\eta$ of the critical correlation
are given by the standard values
\begin{eqnarray}
\nu^{SR}_{MF} && = \frac{1}{2}
\nonumber \\
\eta^{SR}_{MF} && =0
\label{mfsr}
\end{eqnarray}

The upper critical dimension $d_u$ above which mean-field exponents apply
is the value where the mean-field exponents satisfy the hyperscaling relation
\begin{eqnarray}
2-\alpha = (d+z) \nu
\label{hyperscaling}
\end{eqnarray}
leading to
\begin{eqnarray}
d_u=\frac{2-\alpha_{MF}}{\nu^{SR}_{MF}}-z^{SR} =3
\label{dusr}
\end{eqnarray}

\subsection{ Correlation exponents for the long-ranged case  }

For the long-ranged model in dimension $d$
\begin{eqnarray}
J(r) \propto r^{-d-\sigma}
\label{jrd}
\end{eqnarray}
the Gaussian fixed-point of the paramagnetic phase $h>h_c $
corresponds to the correlation \cite{dutta2001}
\begin{eqnarray}
C({\vec r},t) \equiv  <S_{r,t} S_{0,0} > = \int_{-\infty}^{+\infty} \frac{d^d{\vec k}}{(2 \pi)^d}
 \int_{-\infty}^{+\infty} \frac{d\omega}{2 \pi}
\frac{e^{i ({\vec k}.{\vec r} +\omega t)} }{ ( h-h_c ) + \omega^2+\vert {\vec k} \vert^{\sigma}}
\label{correconnected}
\end{eqnarray}
where the singular term $\vert {\vec k} \vert^{\sigma} $ replaces the standard
Short-Ranged term $k^2$ for $\sigma \leq 2$,
with the following consequences :

(i) the dynamical exponent $z$ is fixed by the anisotropy between $\omega^2$
and $\vert {\vec k} \vert^{\sigma} $
\begin{eqnarray}
z^{LR}_{MF} && = \frac{\sigma}{2} 
\label{zmf}
\end{eqnarray}

(ii) the spatial correlation length exponent $\nu$ is fixed by the balance
between the terms $\vert h-h_c \vert $ and $\vert k \vert^{\sigma}$
\begin{eqnarray}
\nu^{LR}_{MF} && =\frac{1}{\sigma} 
\label{numf}
\end{eqnarray}

(ii) the temporal correlation length is fixed by the balance
between the terms $(h-h_c) $ and $\omega^2$, so that the gap 
(corresponding to the inverse temporal correlation length) vanishes as
\begin{eqnarray}
\Delta_{MF} && =  (h-h_c)^{g_{MF}} 
\ \ {\rm with } \ g_{MF}=\frac{1}{2}
\label{gapmf}
\end{eqnarray}
as in the short-ranged case. By consistency with Eq. \ref{zmf} and \ref{numf},
one has of course $g_{MF}=z^{LR}_{MF} \nu^{LR}_{MF}$.

(iv) the global static susceptibility diverges 
\begin{eqnarray}
\chi(h) = \int_{-\infty}^{+\infty} d^d{\vec r}  \int_{-\infty}^{+\infty} dt
C({\vec r},t) = \frac{1 }{ ( h-h_c )^{\gamma_{MF}} }
\ \ {\rm with } \ \gamma_{MF}=1
\label{susceptibility}
\end{eqnarray}

(v) at criticality $h=h_c$, the correlation reads
\begin{eqnarray}
C_{criti}({\vec r},t) &&  =  \int_{-\infty}^{+\infty} \frac{d^d{\vec k}}{(2 \pi)^d}
e^{i {\vec k}.{\vec r}}
 \int_{-\infty}^{+\infty} \frac{d\omega}{2 \pi}
\frac{ e^{i \omega t}  }{  \omega^2+\vert \vec k \vert^{\sigma}}
 = \int_{-\infty}^{+\infty} \frac{d^d{\vec k}}{(2 \pi)^d}
e^{i {\vec k}.{\vec r}}
\frac{ e^{- \vert \vec k \vert^{\frac{\sigma}{2}} t } }{2\vert  \vec k\vert^{\frac{\sigma}{2}} }  
\label{correcriti}
\end{eqnarray}
In particular, the spatial correlation at coinciding points follows the power-law
\begin{eqnarray}
C_{criti}({\vec r},t=0) &&  = \int_{-\infty}^{+\infty} \frac{d^d{\vec k}}{(2 \pi)^d}
\frac{e^{i {\vec k}.{\vec r}}  }{2\vert  \vec k\vert^{\frac{\sigma}{2}} }  
\propto r^{- (d-\frac{\sigma}{2}  )}  \equiv r^{- (d-2+z+\eta)}
\label{correcritit0}
\end{eqnarray}
with
\begin{eqnarray}
\eta^{LR}_{MF} && =2-\sigma 
\label{etamf}
\end{eqnarray}

The upper critical value $\sigma_u$ below which mean-field exponents apply
is the value where the mean-field exponents satisfy the hyperscaling relation
of Eq. \ref{hyperscaling}
\begin{eqnarray}
2-\alpha_{MF} = (d+z^{LR}_{MF}) \nu^{LR}_{MF}
\label{hyperscalinglr}
\end{eqnarray}
leading to \cite{dutta2001}
\begin{eqnarray}
d_u= \frac{3 \sigma_u }{2}
\label{du}
\end{eqnarray}

In particular in dimension $d=1$, the mean-field region $d>d_u$
corresponds to the region $\sigma<\sigma_u$ with
\begin{eqnarray}
\sigma_u= \frac{2}{3}
\label{sigmaupperlr}
\end{eqnarray}
For comparison with the RG procedure described in the text,
it is useful to mention the corresponding exponents
\begin{eqnarray}
\nu (\sigma_u= \frac{2}{3}) && = \frac{1}{\sigma_u} =\frac{3}{2}
\nonumber \\
z (\sigma_u= \frac{2}{3}) &&  = \frac{\sigma_u}{2} = \frac{1}{3}
\label{sigmaunuz}
\end{eqnarray}
and the magnetic exponent
\begin{eqnarray}
x (\sigma_u= \frac{2}{3}) && = \frac{\beta_{MF}}{\nu (\sigma_u= \frac{2}{3})} =\frac{1}{3}
\label{sigmaux}
\end{eqnarray}
Note that at the upper critical value $\sigma_u$, 
one can still use the standard finite-size-scaling valid 
in the non-mean-field region $\sigma>\sigma_u$ with the mean-field exponents valid in the
mean-field region $0<\sigma<\sigma_u$.
But for $0<\sigma<\sigma_u$, the standard finite-size-scaling does not hold anymore
and is replaced by modified finite-size-scaling
(see the series of recent works \cite{bb2012,bb2013,bb2014,bb2015} 
and references therein).

\end{document}